\documentclass[preprint]{JHEP3}

\def\vereq#1#2{%
 \lower3\p@\vbox{%
  \baselineskip1.5\p@
  \lineskip1.5\p@
  \ialign{$\m@th#1\hfill##\hfil$\crcr#2\crcr\sim\crcr}%
 }%
}%

\def\diagram#1{{\normallineskip=8pt
       \normalbaselineskip=0pt \matrix{#1}}}

\def\diagramrightarrow#1#2{\smash{\mathop{\hbox to 
.8in{\rightarrowfill}}
        \limits^{\scriptstyle #1}_{\scriptstyle #2}}}

\def\diagramleftarrow#1#2{\smash{\mathop{\hbox to .8in{\leftarrowfill}}
        \limits^{\scriptstyle #1}_{\scriptstyle #2}}}

\def\diagramdownarrow#1#2{\llap{$\scriptstyle #1$}\left\downarrow
    \vcenter to .6in{}\right.\rlap{$\scriptstyle #2$}}

\title{Conformal Enhancement of Holographic Scaling 
\\
in Black Hole Thermodynamics:
\\
A Near-Horizon Heat-Kernel Framework}

\preprint{\large\sf NSF-KITP-07-152}

\author{
 Horacio E. Camblong\\
 Department of Physics, 
University of San Francisco, 
\\ 
San Francisco, California 94117-1080, U.S.A. 
\\
E-mail:
\sf{camblong@usfca.edu}}

\author{
Carlos R. Ord\'{o}\~{n}ez\\
Department of Physics
\&
\\
World Laboratory Center for Pan-American Collaboration in Science and
Technology, 
\\
University of Houston Center, 
University of Houston,
\\
Houston, Texas 77204-5506, U.S.A.
\\
E-mail: 
\sf{ordonez@uh.edu}}

\preprint{}

\abstract{
 Standard thermodynamic treatments of quantum field theory 
in the presence of black-hole backgrounds reproduce 
the black hole entropy by usually specializing 
to the leading order of the heat-kernel or the high-temperature expansion.
By contrast, this work develops a hybrid framework
centered on geometric spectral asymptotics
whereby these assumptions are shown to be unwarranted
insofar as black hole thermodynamics is concerned. 
The approach---consisting of the concurrent use of 
near-horizon and heat-kernel asymptotic expansions---leads to
a proof of the {\em holographic scaling of the entropy\/}
as a universal feature driven by conformal quantum mechanics. 
}


\begin{document}

\section{Introduction: thermodynamics and heat kernel approach}
\label{sec:intro}

The scaling of the black hole entropy 
with the area of the event horizon has been confirmed by multiple
lines of research~\cite{Bekenstein_entropy,BH-thermo_reviews}.
Its existence and robustness suggest that the relevant gravitational states 
come in a denumerable set associated with a Planck-area partitioned  
horizon~\cite{BH-thermo_reviews,thooft:85,Frolov-Fursaev}.
This is a holographic property:
the horizon encodes information at the quantum level---with
generalizations to a holographic principle~\cite{holographic}
 and the AdS/CFT correspondence~\cite{AdS/CFT}.
Conversely, the near-horizon physics of black holes may provide
guiding hints into the nature of quantum gravity.
As a first ``semiclassical'' approach, 
the quantum fields act as probes of the gravitational background,
and an {\em ultraviolet catastrophe\/}~\cite{BH-thermo_reviews}
affects the spectral and thermodynamic functions.
In its most basic format, the mandatory regularization 
is called the brick-wall model~\cite{thooft:85},
with the dominant part of the entropy
arising from within a Planck-length skin of the horizon---a method
equivalent to the entanglement entropy~\cite{entanglement}.
Most importantly, a universal near-horizon behavior,
known as {\em conformal quantum mechanics\/} (CQM), appears to drive 
the thermodynamics~\cite{BH_thermo_CQM,semiclassical_BH_thermo}.

In this paper we explore the near-horizon emergence
of black hole thermodynamics with the aid of the {\em heat-kernel method\/}. 
Specifically, our framework is based on a direct evaluation 
of the spectral functions to highlight the ultraviolet catastrophe 
and the concomitant role of CQM.
As we show below, the near-horizon CQM contribution to black hole 
thermodynamics trades the orthodox hierarchy in favor of a 
{\em universal holographic scaling\/} that weighs similarly 
all terms in the heat-kernel expansion, 
with the entropy being proportional to the horizon area.
This {\em conformal enhancement\/} 
reveals the holographic nature of the
entropy and entails a revision of semiclassical
treatments of quantum field theory near an event horizon.

The proposed framework is motivated by the need to go
beyond the ``thermodynamic limit'' of regular systems, 
in which boundary contributions are 
neglected compared to the bulk. 
In effect, as boundaries become prominent, e.g., 
for the Casimir effect~\cite{Fulling} or for mesoscopic scales,
boundary corrections are also required. 
In addition, in the presence of a nontrivial bulk geometry,
curvature-dependent corrections set in. 
For black-hole backgrounds, one would expect 
both types of corrections to be relevant {\em a priori\/}.
The ensuing {\em geometric spectral asymptotics\/}
involves a hierarchy of curvature and boundary contributions
organized by the heat kernel
formalism~\cite{Fulling,GR_Hawking,Vassilevich},
which starts with the equation
\begin{equation}
{\mathcal K} \, \Phi  
= 
\lambda \, \Phi
\; ,
\label{eq:Klein-Gordon_curved}
\end{equation}
from which the operator $ {\mathcal K} $
yields the system's spectral functions.
Then, the heat-kernel trace,
$ Y_{\mathcal K}  (\tau)  \equiv 
{\rm Tr} 
\,
\exp 
\left( - \tau {\mathcal K} \right)$,
provides the main asymptotic expansion,
 \begin{equation}
 Y_{\mathcal K}  (\tau) 
 \stackrel{(\tau \rightarrow 0)}{\sim} 
\sum_{j=0}^{\infty}
a_{j} \! \left( {\mathcal K}  \right) \,  \tau^{r_{j} (d)} 
\; ,
\label{eq:Hamidew_expansion}
\end{equation}
where 
$r_{j} (d) =
\left(j - d \right)/2
$
and $d$ is the relevant 
dimensionality~\cite{Fulling,GR_Hawking,Vassilevich}---in 
our canonical framework described below, 
after a dimensional reduction, $d$ will be the spatial dimension.

Our goal is to build the thermodynamics {\em ab initio\/}
and bypass the  assumptions built into the standard thermodynamic 
treatments of quantum field theory in black-hole backgrounds.
In the proposed framework, the density of modes 
$ \rho_{\mathcal K} (\lambda) $
is a primary spectral function for the system's statistical mechanics.
As $Y_{\mathcal K}  (\tau)
=
\int_{0}^{\infty} 
d \lambda
\,
\exp \left( - \tau \lambda \right)
\,
\rho_{\mathcal K}  (\lambda)
$
[for a spectrum in $[0,\infty)$], 
the asymptotic expansion of $ \rho_{\mathcal K} (\lambda)$ 
is the formal inverse Laplace transform 
of eq.~(\ref{eq:Hamidew_expansion}); 
then, the 
{\em spectral counting function\/} 
measuring the cumulative number of 
modes 
\nolinebreak
is
\begin{equation}
N_{\mathcal K} (\lambda)
  \equiv
\int^{\lambda}
d \lambda' \rho_{\mathcal K} (\lambda')
 \stackrel{(\lambda \rightarrow \infty)}{\sim} 
\sum_{j=0}^{\infty}
 \frac{
 a_{j} \! \left( {\mathcal K}  \right)
 }{ \Gamma
\mbox{\large  $\left(  \right. $ } \! \! \!
 - r_{j} (d)  + 1
\! \mbox{\large  $\left.  \right) $ }
}
\,
 \lambda^{-r_{j} (d)}
\; .
\label{eq:Hamidew_expansion_number}
\end{equation}
In particular, with the relevant metric $\tilde{\gamma}_{ij} $  
[to be defined in conjunction with 
eq.~(\ref{eq:Klein-Gordon_curved})],
\begin{equation}
a_{0}
\equiv
 a_{0} \! \left( {\mathcal K}  \right)
 =
\frac{1}{
\left( 4 \pi \right)^{d/2}}
\,
\int_{  {\mathcal M} }
 d^{d} x \,
\sqrt{ \tilde{\gamma} }
\; 
\label{eq:Hamidew_a0_BH}
\end{equation}
yields Weyl's asymptotic formula~\cite{Fulling}
\begin{equation}
N^{(0)}_{\mathcal K} (\lambda)
 \stackrel{(\lambda \rightarrow \infty)}{\sim} 
\frac{ \tilde{\mathcal V}_{d} 
\left(  {\mathcal M}  \right)
}{ 
\left( 4 \pi \right)^{d/2} 
\,
\Gamma 
\left( 1 + d/2 \right) }
\,
\lambda^{d/2}
=
\left( 2 \pi \right)^{-d} 
\,
{\mathcal B}_{d} 
\,
 \tilde{\mathcal V}_{d} \left(  {\mathcal M}  \right)
\,
\lambda^{d/2}
\; ,
\label{eq:Weyl-formula}
\end{equation}
in terms of the manifold volume
$\tilde{\mathcal V}_{d} \left(  {\mathcal M}  \right)$
and the volume ${\mathcal B}_{d} $  of the unit ball.
Equation~(\ref{eq:Weyl-formula})
reproduces Euclidean-space bulk thermodynamics 
and agrees with the mode-counting algorithms of standard 
WKB, phase-space methods, and the brick-wall model. 
In this paper: 
(i) we generalize the latter by going beyond
eq.~(\ref{eq:Weyl-formula}) via the
inclusion of terms with $j \neq 0$;
and 
(ii) show that the orthodox heat-kernel hierarchy 
breaks down in the presence of an event horizon, leading to 
the emergence of universal holographic scaling.
In a sense, this is reminiscent of the metaphor
``hearing the shape of a drum''~\cite{Kac-drum}:
for black holes, when one is effectively
{\em ``probing the event horizon,''\/}
the output is invariably a holographic response.

\section{Quantum field theory: canonical reduction procedure}
\label{sec:QFT}

For our current purposes, we take the broad family 
of generalized Schwarzschild metrics,
\begin{equation}
 ds^{2}
=
- f (r) \,  dt^{2}
+
\left[ f(r) \right]^{-1} \, dr^{2}
+ r^{2} \,
 d \Omega^{2}_{(D-2)}
\;  ,
\label{eq:RN_metric}
\end{equation}
 in $D$ spacetime dimensions, 
as a background probe for quantum fields.
For the sake of simplicity,
we consider one species of a scalar field,
with  action ($D \geq 4$)
\begin{equation}
S
=
-
\frac{1}{2}
\int
d^{D} x
\,
\sqrt{-g}
\,
\left[
g^{\mu \nu}
\,
\nabla_{\mu} \Phi
\,
\nabla_{\nu} \Phi
+
m^{2} \Phi^{2}
+  \xi R \Phi^{2}
\right]
\; ,
\label{eq:scalar_action}
\end{equation}
and
generalizations that will be discussed elsewhere.
We begin our first-principle framework
with a multiple reduction procedure that consists of
the following three steps:

1. The reduction of the original $D$-dimensional
spacetime geometry to a $d$-dimensional
spatial (with $d=D-1$) geometry, along with 
a Fourier resolution of the fields.

2. A conformal transformation of the metric~(\ref{eq:RN_metric}),
\begin{equation}
\tilde{g}_{ \mu \nu }
=
\Omega^{2}
\,
g_{ \mu \nu }
\; ,
\label{eq:optical-metric}
\end{equation}
with the conformal factor
$\Omega^{2} =  1/|g_{00}| = 1/f(r) $.

3. A near-horizon approximation
with respect to the shifted radial coordinate
\begin{equation}
x  
= 
r - r_{+}
\; 
\label{n-h_variable}
\end{equation}
 away from the horizon
${\mathcal H}$ ($ r = r_{+}$),
with the largest root $r_{+}$ of
 $ g^{rr} (r)= f(r) = 0$.

In the first step of the reduction procedure,
a dimensional reduction trades covariance in favor of 
a {\em constructive thermodynamic approach\/} based on 
frequency-mode counting.
The static metric~(\ref{eq:RN_metric}) yields the
Killing vector $\partial_{t}$,
orthogonal to spacelike hypersurfaces that foliate spacetime,
and permits the separation of the coordinate $t$ via
the basis of modes
$\phi_{s,\pm \omega}(t, \vec{x} ) 
=
\phi_{s} (\vec{x} ) e^{ \mp i  \omega t}$
(with $\omega \equiv \omega_{s}$)
satisfying the Lie-derivative equation
${\mathcal L}_{\partial_{t}} \,
\phi_{s,\pm \omega}(t, \vec{x} ) 
=
\mp i \, \omega 
\,
\phi_{s,\pm \omega}(t, \vec{x} ) $
and the Klein-Gordon equation
$
\left[
\Box
-
\left(
m^{2}
+
\xi R
\right)
 \right] 
\phi_{s, \pm \omega}
=
0
$.
Then, in the generalized-Schwarzschild frame, 
the quantum-field-operator expansion
becomes~\cite{BH_thermo_CQM,semiclassical_BH_thermo}
\begin{equation}
 \Phi(t, \vec{x} )
= \sum_{s}
\left[
        a_{s}
\,
\phi_{s} (\vec{x} )
\,
e^{-i\omega t }
        + a^{\dagger}_{s}
\,
\phi^{*}_{s} (\vec{x} )
\,
e^{i\omega t}
        \right]
\; ,
\label{eq:field_Fourier_expansion}
\end{equation}
while
the Klein-Gordon equation takes the form
\begin{equation}
-
{\Delta}_{(\gamma)} \phi
-
2
\,
\gamma^{ij}
\omega_{i}
\partial_{j} \phi
+
\left(
m^{2}
+
\xi  R
\right)
\phi
=
 \frac{\omega^{2}}{f (r) }
\,
\phi
\; ,
\label{eq:Klein_Gordon_reduced_to_spatial}
\end{equation}
where ${\Delta}_{(\gamma)}$ is
the Laplace-Beltrami operator in the spatial metric
$
\gamma_{ij} dx^{i} dx^{j}
=
\left[ f(r) \right]^{-1} \, dr^{2}
+ r^{2} \,
 d \Omega^{2}_{(D-2)}
$ 
and
$
\omega_{i}
=
\frac{1}{2} \,
\partial_{i}
\left(
\ln \sqrt{ |g_{00}|}
\right)
$
is a one-form (from dimensional reduction).

The second step of the reduction procedure
originates in two complicating features
\newline
of the Klein-Gordon equation~(\ref{eq:Klein_Gordon_reduced_to_spatial}):
(i) first-order derivatives;
(ii) the placement of the 
\newpage
\noindent
function $f(r)$ on its right-hand side.
The latter feature suggests the need for a transformation that
recasts eq.~(\ref{eq:Klein_Gordon_reduced_to_spatial})
into the form~(\ref{eq:Klein-Gordon_curved}), with the 
eigenvalue $\lambda $ as function of the
frequency $\omega$ (for the thermodynamic counting of modes).
The resulting conformal transformation~(\ref{eq:optical-metric})
of eq.~(\ref{eq:RN_metric}) defines the spacetime
{\em optical metric\/}~\cite{optical_metric,deAlwis-Ohta,Barbon-Emparan},
whose spatial part 
$\tilde{\gamma}_{ij}
=
\left[ f (r) \right]^{-1}
\, \gamma_{ij}
$
[where $g_{0 0 } = - f(r)$ for the class of geometries
of eq.~(\ref{eq:RN_metric})]
will be used as the relevant spatial metric
[cf. eq.~(\ref{eq:Hamidew_a0_BH})]
and the tilde denotes 
all geometric quantities defined therewith.
As a result, in eq.~(\ref{eq:Klein-Gordon_curved}),
the differential operator becomes
\begin{equation}
{\mathcal K}
=
- \left(
\tilde{\gamma}^{ij} \tilde{D}_{i} \tilde{D}_{j}
+
\tilde{E}
\right)
\;
\label{eq:diff-op-BH_normal-form}
\end{equation}
and the eigenvalue 
is seen to be
$
\lambda
=
\omega^{2}
$.
In this approach, 
the operator~(\ref{eq:diff-op-BH_normal-form})
is built from the generalized covariant derivatives
$\tilde{D}_{j}
=
\tilde{\nabla}_{j}
+
\tilde{\omega}_{j}
$,
with 
\begin{equation}
\tilde{\omega}_{i}
=
\frac{ (d-1)}{2}
 \, \partial_{i}
\left(
\ln \sqrt{ |g_{00}|}
\right)
\; 
\label{eq:optical_one-form}
\end{equation} 
obtained from the first-order terms
in the conformally transformed 
Klein-Gordon equation. 
Incidentally, the sequence of 
rearrangements of eq.~(\ref{eq:Klein_Gordon_reduced_to_spatial}),
leading to eq.~(\ref{eq:diff-op-BH_normal-form})
through the generalized covariant derivatives
$\tilde{D}_{j}$,
 is equivalent to the Liouville transformations~\cite{forsyth:Liouville}
of refs.~\cite{BH_thermo_CQM,semiclassical_BH_thermo},
thus solving the problem of transforming away the first-order derivatives.
The one-form $\tilde{\omega}_{i}$
of eq.~(\ref{eq:optical_one-form})
can be reinterpreted as defining an 
additional vector-fiber structure~\cite{Vassilevich}
and plays the role of a connection.
Moreover,
 eq.~(\ref{eq:optical_one-form})
shows it is an exact form, 
so that its associated curvature form $\tilde{\Omega}_{ij}$
vanishes---thus,
the latter is absent from the heat-kernel expansion.
Finally, the normal invariant or independent-term scalar
in eq.~(\ref{eq:diff-op-BH_normal-form}),
\begin{equation}
\tilde{E}
\equiv
\tilde{E} (r)
=
-
f (r)
\,
\left(
m^{2}
+
\xi  R
\right)
-
\tilde{\gamma}^{ij}
\left(
\partial_{i} \tilde{\omega}_{j}
-
\tilde{\omega}_{ l }
\tilde{\Gamma}_{ ij }^{ l }
+
\tilde{\omega}_{i} \tilde{\omega}_{j}
\right)
\; ,
\label{eq:normal_invariant_modified-RN-metric}
\end{equation}
includes ``extra terms'' 
from the rearrangement of the generalized covariant derivative 
$\tilde{D}_{j}$.

The third step in the reduction procedure involves a number of subtleties
that will be discussed in the next section.

\section{Near-horizon heat-kernel approach}
\label{sec:heat-kernel}

In this section we elaborate upon the two major
technical building blocks
of our {\em near-horizon spectral asymptotics\/}:
the computation of the heat-kernel coefficients 
and the near-horizon approximation.
Specifically,
our hybrid framework involves
a number of subtleties arising from the simultaneous 
use of the above ingredients.

Firstly,
the computation of the heat-kernel or HaMiDeW coefficients involves 
the geometrically structured integrals~\cite{Fulling,Vassilevich}
\begin{equation}
a_{j}
\equiv
a_{j} \left( {\mathcal K}  \right)
 =
\left\{
\begin{array}{ll}
\left( 4 \pi \right)^{-d/2}
\,
\left[
\int_{  {\mathcal M} }
 d^{d} x \,
\sqrt{ \tilde{\gamma} }
\;
\Gamma_{j}^{(\mathcal M)}
+
\int_{  { {\partial \mathcal M} } }
 d^{d-1} x \,
\sqrt{ \tilde{h} }
\;
\Gamma_{j}^{(\partial \mathcal M)}
\right]
&
{\rm if} \; $j$ \; {\rm is \; even}
\; ,
\\
\left( 4 \pi
 \right)^{-(d-1)/2}
\,
\int_{  { {\partial \mathcal M} } }
 d^{d-1} x \,
\sqrt{ \tilde{h} }
\;
\Gamma_{j}^{(\partial \mathcal M)}
&
{\rm if} \; $j$ \; {\rm is \; odd}
\; ,
\label{eq:Hamidew_aj_BH}
\end{array}
\right.
\end{equation}
which generalize eq.~(\ref{eq:Hamidew_a0_BH})
and lead to extensions of Weyl's asymptotic formula~(\ref{eq:Weyl-formula})
through eq.~(\ref{eq:Hamidew_expansion_number}).
The structural form and geometric content of eq.~(\ref{eq:Hamidew_aj_BH})
are uniquely determined by the optical metric~(\ref{eq:optical-metric})
and the operator (\ref{eq:diff-op-BH_normal-form}): the integrands
$\Gamma_{j}^{(\mathcal M)}$ and $\Gamma_{j}^{(\partial \mathcal M)}$
are additively built from the primitive ``HaMiDeW  invariants,'' 
with appropriate combinatoric coefficients.
In turn, the invariants are multiplicatively
assembled, in all allowed geometric combinations
at a given order, from the
``geometrical building blocks''~\cite{Fulling,Vassilevich}:
(i) the Riemann tensor
 $
\tilde{R}^{ij}_{\; \; \, kl}
$;
(ii)
the normal invariant
$\tilde{E}$;
and (iii)
the extrinsic curvature
$
\tilde{K}^{a}_{\; \,  b}
$
(with the curvature form $\tilde{\Omega}_{ij}$ being
absent herein due to its identically vanishing value).
In essence, from a given geometry,
the ``HaMiDeW
operation'' amounts to 
generating the set of coefficients~(\ref{eq:Hamidew_aj_BH}), 
i.e., the formal operator
$\mathfrak{H}
\left[
{\mathcal K}  
\right]
 =
\left\{
 a_{j} \! \left( {\mathcal K}  \right)
  \right\}_{j \geq 0}
$,
whence the asymptotics of the spectral 
functions can be computed, e.g.,
$
 N_{\mathcal K} (\lambda)
$
in eq.~(\ref{eq:Hamidew_expansion_number}).
This operation is displayed
by the horizontal arrows in 
the diagram~(\ref{eq:diagram_H-nh}).

In addition, it should be noticed that, 
for the evaluation of eq.~(\ref{eq:Hamidew_aj_BH}),
the boundary is the event horizon---in this spatially reduced view,
the horizon ${\mathcal H\/}$ represents the {\em spatial boundary\/} of the 
region classically accessible to an external observer.
Thus, in a canonical approach with frequency-dependent spectral functions,
the possible inclusion of terms arising from the 
spatial boundary ${\mathcal H}$ cannot be 
ignored.
%---contrary to standard practice in applications of the brick-wall model.

Secondly, the other technical ingredient
in our approach---which constitutes the
third step of the reduction procedure of the
previous section---amounts 
to isolating the dominant physics as $ r \sim r_{+}$, i.e.,
performing an expansion in the near-horizon coordinate $x$ 
of eq.~(\ref{n-h_variable}) and abstracting the leading parts.
This process, which will be represented by the symbol 
$\stackrel{(\mathcal H)}{\sim}$,
is displayed by the vertical arrows in 
the diagram~(\ref{eq:diagram_H-nh}).
Moreover, 
the near-horizon hierarchical scheme
serves a twofold purpose:
(a)
displaying the emergence of CQM as 
the leading physics of the modes that generate the 
field operator~(\ref{eq:field_Fourier_expansion});
(b)
 providing a consistent 
scaling of all the heat-kernel coefficients,
which ultimately causes the holographic nature of the entropy.
Therefore, this ingredient 
uncovers, {\em inter alia\/},
that CQM is responsible for the 
{\em universal holographic scaling\/} of spectral functions and the entropy.
The resultant emergent behavior will be derived in the next section.

Having defined these two technical steps as formal operations,
the question naturally arises as to
%their sequential order and 
the legitimacy of combining
them into a unified approach and the order 
of their sequential application.
More precisely, one may ask whether the diagram
\begin{equation}
\diagram{
\fbox{
$
\begin{array}{c}
 \mbox{{\sf GLOBAL}}
\\
\mbox{ {\sf GEOMETRY} }
 \end{array}
$ }
  &  \diagramrightarrow{ 
\mbox{
$\mathfrak{H}$
}
}{}
          &
\mathfrak{H}
\left[
{\mathcal K}  
\right]
 =
\left\{
 a_{j} \! \left( {\mathcal K}  \right)
  \right\}_{j \geq 0}
  \cr
 \diagramdownarrow{ \mbox{$
 \stackrel{(\mathcal H)}{\sim}         
                  $
                   } }{}  & &
 \diagramdownarrow{ \mbox{
$
 \stackrel{(\mathcal H)}{\sim}
$
            } }{}  \cr
\fbox{
$
\begin{array}{c}
\mbox { {\sf NEAR-HORIZON} }
\\
\mbox{  {\sf GEOMETRY} }
 \end{array}
$ }
&
  \diagramrightarrow{ \mbox{
$\mathfrak{H}$
  }  }{}
    &  
\mathfrak{H}^{ (\mathcal H) }
\left[
{\mathcal K}  
\right]
 =
\left\{
 a_{j}^{ (\mathcal H) } \! \left( {\mathcal K}  \right)
  \right\}_{j \geq 0}
}
\;  \;  
\label{eq:diagram_H-nh}
\end{equation}
is indeed commutative.
In essence, the legitimacy of 
the ordering consisting of the application
of $\mathfrak{H}$ followed by the near-horizon approximation
$\stackrel{(\mathcal H)}{\sim}$ 
is validated {\em a priori\/} by the large body of well-established results 
on the heat-kernel approach
for compact manifolds~\cite{Fulling}.
Thus, the actual computations can be performed by introducing an infrared
cutoff---indeed, this has been a standard procedure in the 
literature, as required by the usual statistical thermodynamic 
arguments (e.g., ref.~\cite{thooft:85}).
When this is done,
all the bulk integrals
in eq.~(\ref{eq:Hamidew_aj_BH}) are finite, except for 
possible divergences caused by the behavior of the metric
upon crossing the horizon. 
However, the latter divergences (which amount to the 
``ultraviolet catastrophe'' mentioned in section~\ref{sec:intro})
can be regularized via a near-horizon brick-wall cutoff $a$,
as further discussed and illustrated in the next section.
Consequently, with the ultraviolet regulator $a$
and an infrared thermodynamic cutoff, 
all the coefficients~(\ref{eq:Hamidew_aj_BH}) are strictly
finite.

As a result, the near-horizon expansion of the heat-kernel coefficients is 
justified as follows. Each coefficient involves integrals
of the form $\int d \mu \, \Gamma$, which are hierarchically
organized in powers of $x \equiv a$.
 The invariant integrands are built from the HaMiDeW invariants
defined above, which are all finite at the horizon, 
even in the absence of a regulator---a result that is corroborated in
the next section. By contrast, the measures $d \mu$ (both for the
bulk and boundary contributions) are divergent and provide the leading 
scaling with respect to $a$;  then,
in the Landau big-O notation:
 $\Gamma = O(a^{p})$, with $p \geq 0$,
and
$d \mu  = O(a^{q})$,
with $q < 0$.
The ensuing hierarchical
process is used for the selection of the dominant order.
Due to the finiteness of all expressions thus regularized, this 
process can be systematically applied to all the 
integrals and provides a result that
is independent of the 
ordering in the diagram~(\ref{eq:diagram_H-nh}).
The reversal of this ordering, with the
near-horizon operation
$\stackrel{(\mathcal H)}{\sim}$ 
preceding the application
of $\mathfrak{H}$,
is explicitly used in the next section.
Parenthetically,
 a more detailed evaluation of the global coefficients will
be discussed in a forthcoming paper, along with explicit
generalized expressions for the HaMiDeW invariants.

\section{Near-horizon physics: building blocks \& thermodynamics}
\label{sec:n-h&thermo}

In this section we display the emergence of CQM and the thermodynamics 
via the near-horizon heat-kernel formalism of the previous section.

 The conformal behavior of the leading near-horizon physics 
is exposed from eq.~(\ref{eq:diff-op-BH_normal-form}) 
when $ r \sim r_{+}$.
We will only consider the {\em nonextremal\/} case,
for which
$f'_{+}
\equiv f'(r_{+})
\neq 0
$
is related to the surface gravity 
$
\kappa
=
 f'_{+} /2$
and interpreted with
the physics of CQM~\cite{BH_thermo_CQM,semiclassical_BH_thermo},
as shown below.
Then, starting from
 $f(r)
 \stackrel{(\mathcal H)}{\sim} 
f'_{+}  \, x
$, 
the metric coefficients are
$g_{00}  \stackrel{(\mathcal H)}{\sim} - f_{+}' x$
and $g_{rr}  \stackrel{(\mathcal H)}{\sim} 1/\left( f_{+}' x \right)$;
thus,
the radial and angular optical counterparts
are  $\tilde{\gamma}_{rr} 
 \stackrel{(\mathcal H)}{\sim} 
1/ \left(  2 \kappa x \right)^{2}$
and 
 $\tilde{\gamma}_{ab} 
 \stackrel{(\mathcal H)}{\sim} 
{\gamma}_{ab} 
/  \left( 2 \kappa x \right)
$,
while the Christoffel symbols are, e.g., 
$\tilde{\Gamma}^{r}_{rr}  \stackrel{(\mathcal H)}{\sim}  - 1/x$
and 
$\tilde{\Gamma}^{r}_{ab}  \stackrel{(\mathcal H)}{\sim}  
2 \kappa^{2} \, \tilde{\gamma}_{ab} \, x$.
In addition, the one-form~(\ref{eq:optical_one-form})
turns into
\begin{equation}
\tilde{\omega}_{i}
 \stackrel{(\mathcal H)}{\sim} 
\frac{ (d-1)}{4}
\,
\frac{1}{x}
\,
 \, \delta^{r}_{i}
\label{eq:n-h_optical_one-form}
\end{equation}
while the normal invariant~(\ref{eq:normal_invariant_modified-RN-metric}),
governed by the ``extra terms,'' becomes
\begin{equation}
\tilde{E} 
 \stackrel{(\mathcal H)}{\sim} 
\frac{ \left( d -1 \right)^{2} }{ 4 } 
\, 
\kappa^{2}
\; .
\label{eq:n-h_normal-invariant}
\end{equation}
As a result, 
the rescaled leading form of the operator~(\ref{eq:diff-op-BH_normal-form}) 
becomes
\begin{equation}
-\frac{{\mathcal K}}{ 4 \kappa^{2} }
 \stackrel{(\mathcal H)}{\sim}  
x^{2}  
\,
\left[  \partial_{x} - \frac{ \left( d -5 \right)}{4} \frac{1}{x} \right]
\,
\left[  \partial_{x} + \frac{ \left( d -1 \right)}{4} \frac{1}{x} \right]
+
\frac{ \left( d -1 \right)^{2} }{ 16 } 
 =  
 x^{2} \partial_{x}^{2}
+
x \partial_{x} 
\; ,
\label{eq:n-h_diff-op-BH_normal-form}
\end{equation}
which is manifestly {\em homogeneous of degree zero\/}.
The conformal invariance 
of eq.~(\ref{eq:n-h_diff-op-BH_normal-form}) arises from the
uniform degree of homogeneity
of the kinetic and ``extra terms,''
while the surface gravity provides a scale for comparison with the
eigenvalues of eq.~(\ref{eq:Klein-Gordon_curved}), 
$\lambda = \omega^{2}$.
Therefore, the {\em optical-metric version of near-horizon CQM\/}
is given by eq.~(\ref{eq:Klein-Gordon_curved}), 
with the operator~(\ref{eq:n-h_diff-op-BH_normal-form});
in addition, via Liouville transformations~\cite{forsyth:Liouville},
this is equivalent to the other two forms of near-horizon CQM
derived in ref.~\cite{semiclassical_BH_thermo}
(for example, the conformal coupling is $\Theta = \omega/f_{+}'$).

Most importantly, a concurrent near-horizon expansion extracts 
the leading behavior in
$x$. This is unlike the orthodox procedure, 
in which the geometrical building blocks are evaluated
from their global definitions 
over the manifold and its boundary.
The first building block, the Riemann tensor,
 has a near-horizon {\em maximally symmetric\/} structure
\begin{equation}
\tilde{R}^{ij}_{\; \; \, kl}
 \stackrel{(\mathcal H)}{\sim} 
- \kappa^{2}
\left( 
\delta^{i}_{k} \delta^{j}_{l}
-
\delta^{i}_{l} \delta^{j}_{k}
\right)
\; ,
\label{eq:n-h_Riemann-tensor}
\end{equation}
with its counterpart
$\tilde{R}_{ij kl}
 \stackrel{(\mathcal H)}{\sim} 
- \kappa^{2}
\left( 
g_{ik} g_{jl}
-
g_{il} g_{jk}
\right)$.
This symmetry is due to the 
single-parameter characterization of the near-horizon geometry
by the {\em surface gravity\/} $\kappa$; 
i.e.,
$\kappa$ sets uniquely the natural horizon scale
of all physical quantities.
The corresponding Ricci tensor
$
\tilde{R}^{i}_{\; j}
 \stackrel{(\mathcal H)}{\sim} 
- \left( d - 1 \right)
\,
\kappa^{2}
\,
\delta^{i}_{j} 
$
and scalar curvature
$
\tilde{R}
 \stackrel{(\mathcal H)}{\sim} 
- d \left( d - 1 \right)
\,
\kappa^{2}
$
reflect the same symmetry; and
as a result, contractions of multiple products of the three above 
Riemann-related quantities yield straightforward products
and contractions of the corresponding Kronecker
deltas and metric tensors.
The second building block, 
the extrinsic curvature at the horizon,
\begin{equation}
\tilde{K}^{a}_{\; \,  b}
 \stackrel{(\mathcal H)}{=} 
 \kappa
\,
\delta^{a}_{\, b} 
\; ,
\label{eq:n-h_extrinsic-curvature}
\end{equation}
is also maximally symmetric due to 
the geometrically determining role of $\kappa$.
Correspondingly, it generates the following contractions:
$
\tilde{K}
 \equiv 
\tilde{K}^{a}_{\; \,  a}
 \stackrel{(\mathcal H)}{=} 
 \left( d -1 \right)
\,
\kappa
$,
$\tilde{K}^{a}_{\; \,  b}
\,
\tilde{K}^{b}_{\; \,  a}
  \stackrel{(\mathcal H)}{=} 
 \left( d -1 \right)
\,
\kappa^{2}
$,
and so on.
The third building block, 
the normal invariant, displays the near-horizon leading 
behavior~(\ref{eq:n-h_normal-invariant})---also parametrized with $\kappa$.
Therefore, from
Eqs.~(\ref{eq:n-h_normal-invariant}),
(\ref{eq:n-h_Riemann-tensor}), and
(\ref{eq:n-h_extrinsic-curvature}),
 one infers the collective scaling property, 
with respect to the optical metric: 
$\left( {\rm ``Building \; Block''} \right)
 \stackrel{(\mathcal H)}{\sim} 
\kappa^{p} \, \times \, O(x^{0})
$,
where $\kappa^{p}$ is solely determined by dimensional analysis
from the scale $\kappa$.
Thus, the HaMiDeW invariants and the invariant integrands inherit 
the same {\em near-horizon scaling\/} because they are assembled 
from the allowed contracted products of the building blocks
(with the chosen placement of indices in
$
\tilde{R}^{ij}_{\; \; \, kl}
$
and 
$
\tilde{K}^{a}_{\; \,  b}
$);
this results in the naive scaling 
\begin{equation}
\hat{\Gamma}_{j}^{(\mathcal M)}
\equiv
\frac{ \Gamma_{j}^{(\mathcal M)} }{
\kappa^{j}}
 \stackrel{(\mathcal H)}{\sim} 
 O(x^{0})
\; \; ,
\hat{\Gamma}_{j}^{(\mathcal \partial M)}
\equiv
\frac{ \Gamma_{j}^{( \partial \mathcal M)} }{
\kappa^{j-1} }
 \stackrel{(\mathcal H)}{\sim} 
 O(x^{0})
\; .
\label{eq:Hamidew_scaling_integrands}
\end{equation}
As a corollary to
eq.~(\ref{eq:Hamidew_scaling_integrands}),
the near-horizon behavior of the heat-kernel 
coefficients~(\ref{eq:Hamidew_aj_BH})
is completely governed by the scaling of the integral measures:
that of the bulk integral,
\begin{eqnarray}
\tilde{\mathcal V}_{d} \left(  {\mathcal M}  \right)
& \equiv &
\int_{  {\mathcal M} }
 d^{d} x \,
\sqrt{ \tilde{\gamma}} 
\nonumber
\\
& \stackrel{(\mathcal H)}{\sim} &
\frac{ {\mathcal A}_{d-1} }{ 
\left( 2 \kappa \right)^{(d+1)/2} }
\,
\int_{a} \frac{d x}{ x^{(d+1)/2} }
\stackrel{(\mathcal H)}{\sim} 
\frac{ 1 }{ \left( d -1 \right) \, \kappa^{(d+1)/2}  }
\,
2^{- (d-1)/2}
 \,
{\mathcal A}_{d-1}
\,
a^{-(d-1)/2}  
\; ,
\label{eq:nh_bulk-measure}
\end{eqnarray}
known as the ``volume of optical space''~\cite{deAlwis-Ohta},
and that of the boundary integral,
\begin{eqnarray}
\tilde{\mathcal V}_{d-1} \left(  \partial {\mathcal M}  \right)
& \equiv &
\int_{  { {\partial \mathcal M} } }
 d^{d-1} x \,
\sqrt{ \tilde{h} }
\nonumber
\\
& \stackrel{(\mathcal H)}{\sim} &
\frac{ 1 }{  \kappa^{(d-1)/2} } 
\,
2^{- (d-1)/2}
 {\mathcal A}_{d-1}  
\,
a^{-(d-1)/2}  
\;  ,
\label{eq:nh_boundary-measure}
\end{eqnarray}
which are both regularized 
by means of the near-horizon radial-coordinate cutoff $a$.
Because of the mandatory cutoff, the boundary ${\partial \mathcal M}$
is the horizon lifted an amount $a$ in
the radial coordinate $r$: this is the celebrated brick wall
$ {\partial \mathcal M}
=
{\mathcal H\/}_{a} $.
Furthermore, the regularized integrals in 
Eqs.~(\ref{eq:nh_bulk-measure}) and (\ref{eq:nh_boundary-measure})
can be rendered into the invariant forms
\begin{equation}
\tilde{\mathcal V}_{D-1} \left(  {\mathcal M}  \right)
\stackrel{(\mathcal H)}{\sim}
\frac{ 4 }{ \left( D -2 \right) }
\,
\left[ h_{D} \right]^{- (D-2)}
\,
\frac{ {\mathcal A}_{D-2} }{4}
\,
\kappa^{-(D-1)} 
\;
\label{eq:nh_bulk-measure_invariant}
\end{equation}
and
\begin{equation}
\tilde{\mathcal V}_{D-2} \left(  \partial {\mathcal M}  \right)
\stackrel{(\mathcal H)}{\sim}
4
\,
\left[ h_{D} \right]^{- (D-2)}
\,
\frac{ {\mathcal A}_{D-2} }{4}
\,
\kappa^{-(D-2)}  
\;  ,
\label{eq:nh_boundary-measure_invariant}
\end{equation}
via the geometrical radial distance 
$h_{D}
\stackrel{(\mathcal H)}{\sim}
2 \, \sqrt{a/ f_{+}'}
=
\sqrt{2 a/ \kappa}
$
and the horizon area $ {\mathcal A}_{D-2}$ 
in $D$ spacetime dimensions.
In addition, for the remainder of the paper,
we will use the spacetime dimensionality $D  = d + 1$.
Then, Eqs.~(\ref{eq:Hamidew_aj_BH}), 
(\ref{eq:nh_bulk-measure_invariant}), and (\ref{eq:nh_boundary-measure_invariant})
imply that the dimensionless coefficients
$
\hat{a}_{j}
\equiv
a_{j} \,  \kappa^{D-j-1} 
$
are given by
\begin{equation}
\hat{a}_{j}
=
\frac{ 2^{3-D}}{ \pi^{(D-1)/2}  }
\,
\hat{I}_{D}^{(j)}
\,
\left[ h_{D} \right]^{- (D-2)}
\,
\frac{ {\mathcal A}_{D-2} }{4}
\; ,
\label{eq:nh-BH_even-Hamidew-coeffs}
\end{equation}
where
\begin{equation}
\hat{I}_{D}^{(j)}
=
\left\{
\begin{array}{cc}
\frac{1 }{ (D-2)} \,
\hat{\Gamma}_{j}^{(\mathcal M)} 
+
\hat{\Gamma}_{j}^{(\partial \mathcal M)}
&
{\rm if} \; $j$ \; {\rm is \; even}
\; ,
\\
\sqrt{4 \pi} 
\,
\hat{\Gamma}_{j}^{(\partial \mathcal M)}
&
{\rm if} \; $j$ \; {\rm is \; odd}
\; .
\end{array}
\right.
\label{eq:BH-CQM_measure-coeff}
\end{equation}

The main physical quantity of interest in this paper is the entropy,
which we derive through the expansion~(\ref{eq:Hamidew_expansion_number})
of the spectral function,
within the spectral rule~\cite{BH_thermo_CQM,semiclassical_BH_thermo}
\begin{equation}
S
=
- \int_{0}^{\infty}
d \omega  
\,
\ln (1 - e^{-\beta \omega})
\,
\left[
\left(
\omega \frac{d}{d \omega} + 2
\right)
\frac{dN(\omega)}{d\omega}
\right]
\; .
\label{eq:entropy_formula}
\end{equation}
Then, evaluating the dimensionless integrals 
in terms of the Riemann zeta function $\zeta_{R} (z)$
and enforcing the inverse Hawking temperature 
$\beta \equiv \beta_{H} = 2 \pi/\kappa $,
the leading near-horizon part of the
entropy takes the form
\begin{equation}
S
\stackrel{(\mathcal H)}{\sim}
\sum_{j=0}^{\infty}
\frac{
\left( D-j -1 \right) \,
\left( D-j \right)
}{
(2\pi)^{D-j-1}
}
\,
\frac{ \zeta_{R} (D-j) \, \Gamma (D-j-1)
}{
 \Gamma 
\mbox{\large  $\left(  \right. $ } \! \! \! \!
 (D-j + 1 )/2 
\!  \mbox{\large  $\left. \right) $ }
}
\;
\hat{a}_{j} 
\; .
\label{eq:entropy_series}
\end{equation}
The remarkable conclusion of this analysis
is that {\em all the terms in the HaMiDeW 
near-horizon expansion of the entropy,
$S^{(j)}$,
are structurally similar,
being governed by CQM and generating the holographic property\/}
$S \propto {\mathcal A}_{D-2}$.
Specifically,
\begin{equation}
S^{(j)} 
=
c_{D}^{(j)}
\,
\hat{I}_{D}^{(j)}
\left[ h_{D} \right]^{-(D-2)}
\,
\frac{ {\mathcal A}_{D-2} }{4}
\; ,
\label{eq:entropy_order-j}
\end{equation}
which involves the dimensionless prefactors
given by $\hat{I}_{D}^{(j)}$
and
\begin{equation}
c_{D}^{(j)}
 = 
\frac{ 2^{4-D}}{  \pi^{3D/2 - 1 -j} }
\,
\Gamma \left[ \frac{\left( D-j \right)}{ 2} + 1 \right]
\,
\zeta_{R} \left( D -j \right)
\; .
\end{equation}
Incidentally, even though the HaMiDeW  numerical coefficients are 
of order unity, they are also functions of the dimensionality $D$;
thus, they could be zero for particular values of $D$.
For example, the lowest-order dimensionless
 invariant integrands of the heat kernel coefficients---evaluated 
with the usual {\em Dirichlet boundary condition at the brick wall\/}---are:
$
\hat{\Gamma}^{( {\mathcal M} )}_{0}
 \stackrel{(\mathcal H)}{\sim} 
1
$,
$
\hat{\Gamma}^{( \partial {\mathcal M} )}_{1}
 \stackrel{(\mathcal H)}{\sim} 
- 1/4
$,
$
\hat{\Gamma}^{( {\mathcal M} )}_{2}
 \stackrel{(\mathcal H)}{\sim} 
 (D-2) (D-4)/12
$,
$
\hat{\Gamma}^{( \partial {\mathcal M} )}_{2}
 \stackrel{(\mathcal H)}{\sim} 
\left( D - 2 \right)/3  
$,
and so on.
Equation~(\ref{eq:entropy_formula}) suggests that
the terms $j \leq D-1$ are relevant, while 
the higher orders require further analysis.
The ensuing pattern of cancellations and the significance 
of higher orders for different dimensionalities 
will be systematically studied elsewhere.

\section{Context and conclusions}
\label{sec:conclusions}

A more detailed comparison of our near-horizon heat-kernel framework
with the existing literature is in order.
An important point of contact is established by exploring
the temperature expansion for miscellaneous thermodynamic functions.
This can be developed in the
{\em ab initio\/} canonical framework 
from the series~(\ref{eq:Hamidew_expansion_number})
with the spectral counting function $N(\omega)$, via
the counterpart of
eq.~(\ref{eq:entropy_formula}) for a generic thermodynamic function, i.e.,
\begin{equation}
{\mathcal T}
=
- \int_{0}^{\infty}
d \omega
\,
\ln (1 - e^{-\beta \omega})
\,
\hat{\mathcal T}_{\omega} (\beta)
\!
\left[
\frac{ dN(\omega) }{ d\omega }
\right]
\; ,
\label{eq:generic_thermo-function}
\end{equation}
where the operator 
$\hat{\mathcal T}_{\omega}(\beta)$
(differential with respect to $\omega$ and possibly 
$\beta$-dependent)
is function-specific.
Most importantly, for the free energy ${\mathcal T} = F$, 
$\hat{\mathcal T}_{\omega}= - 1/\beta$;
and for the entropy
${\mathcal T} = S$, 
$\hat{\mathcal T}_{\omega}=   2 + \omega \, d/d \omega$.
Substituting eq.~(\ref{eq:Hamidew_expansion_number})
in eq.~(\ref{eq:generic_thermo-function}),
one may write
${\mathcal T} = \sum_{j =0}^{\infty} {\mathcal T}^{(j)}$
[with $j$ being the order in the heat-kernel
series~(\ref{eq:Hamidew_expansion_number})], which
yields a temperature expansion for all thermodynamic functions.
The scaling of these functions, order by order,
follows by using the same approach as that
 leading to the scaling of eq.~(\ref{eq:entropy_formula})---the 
only caveat is that, for comparison purposes, 
here we will not assume that the temperature is the Hawking temperature yet.
From eq.~(\ref{eq:generic_thermo-function}), 
$\mathcal{T}^{(j)} 
\propto
\beta^{ 2 r_{j} (d)  + {\rm deg}_{\beta} ( \hat{\mathcal T}_{\omega} ) }
$,
with ${\rm deg}_{\beta} ( \hat{\mathcal T}_{\omega} ) $ being the degree 
of homogeneity 
of the operator $\hat{\mathcal T}_{\omega} $ with respect to $\beta$.
For example, 
$S^{(j)} \propto
\beta^{ 2 r_{j} (d)}
=
T^{   D - 1 - j }
$ 
for the entropy 
[i.e., $S^{(j)} $ is
eq.~(\ref{eq:entropy_order-j})
times $ \left( \beta \kappa/2 \pi \right)^{ 2 r_{j} (d)} $]
while
$F^{(j)} \propto \beta^{ 2 r_{j} (d) - 1 }
=
T^{   D  - j }
$
for the free energy; more precisely.
\begin{equation}
- F^{(j)} 
\stackrel{(\mathcal H)}{\sim} 
\frac{1}{ \left( D -j \right) \, \beta } \, S^{(j)} 
\stackrel{(\mathcal H)}{\sim} 
\frac{ c_{D}^{(j)} }{ \left( D - j \right)}
\,
\hat{I}_{D}^{(j)}
\left[ h_{D} \right]^{-(D-2)}
\,
\frac{ {\mathcal A}_{D-2} }{4}
\,
\left( \frac{\kappa}{2 \pi } \right)^{ j -D + 1 }
\,
\beta^{ j - D }
\; .
\label{eq:free-energy_order-j}
\end{equation}
In particular, the leading order, $j = 0$:
(i) is equivalent (by definition)
to the thermodynamics based on Weyl's formula~(\ref{eq:Weyl-formula});
(ii)
is seen to reproduce the standard brick-wall results
of refs.~\cite{thooft:85}
and \cite{Frolov-Fursaev}.
Moreover, this a term that is typically singled out in 
refs.~\cite{deAlwis-Ohta,Barbon-Emparan,susskind_uglum,renormalization_Newton_confirmed}
[for example, 
eq.~(3.26) in ref.~\cite{deAlwis-Ohta}, as can be verified 
from our Eqs.~(\ref{eq:nh_bulk-measure_invariant}) and
(\ref{eq:free-energy_order-j})].

Our work shows that usage of the high-temperature limit or equivalent
for black hole thermodynamics may be unwarranted because the 
Hawking temperature is of order unity in surface-gravity units.
Once the comparison above is made, one 
may ask the extent to which these results
modify the standard folklore for the thermodynamics
in the presence of horizons.
The key lesson from this computation
is the realization that, conceptually, the scaling of all orders 
of the HamiDeW 
expansion is identical
when the Hawking temperature is enforced;
thus, by the same token, 
this implies in the end
that the holographic scaling of the entropy
remains intact.
In a sense, the holographic outcome is enhanced---in the language we have
often used in this paper.
In the brick-wall approach,
this specifically yields a correction 
for the expression of the brick-wall elevation, with the main 
conclusions of refs.~\cite{thooft:85}
and \cite{Frolov-Fursaev}
remaining intact.
Possible additional implications of these differences are
discussed next.

In conclusion, we have developed a near-horizon heat-kernel framework
that comprehensively addresses the derivation of black hole thermodynamics 
from conformal quantum mechanics. 
The guiding strategy has been the inclusion of all contributions 
to the density of modes within a brick-wall type model. 
In essence, this hybrid framework combines the conformal properties 
of the horizon with the insight afforded 
by geometric spectral asymptotics; such
format transcends the traditional heat-kernel
hierarchy and the standard brick-wall model.
As an outcome of these extensions, we have found that 
all orders of the heat-kernel expansion and associated spectral functions, 
e.g., the density of modes, exhibit the same scaling behavior 
with respect to the near-horizon expansion. 
At the level of the entropy, the ensuing conformal enhancement
further stresses its holographic nature.
In addition, as these findings entail possible modifications of 
the usual brick-wall type calculations,
they should be included in any discussions of 
the ``species problem''~\cite{thooft:85,BH-thermo_reviews,Frolov-Fursaev}
and the renormalization of Newton's gravitational 
constant~\cite{deAlwis-Ohta,Barbon-Emparan,susskind_uglum,renormalization_Newton_confirmed}.
Moreover, the robustness of holographic scaling and the crucial role 
of scaling and conformal properties for black hole thermodynamics 
call for a deeper interpretation, possibly related to 
conformal field theory~\cite{carlip:near_horizon,solodukhin:99}.
These unresolved issues, as well as an extension of scaling
properties via path integrals, including a detailed comparison with 
refs.~\cite{optical_metric,deAlwis-Ohta,Barbon-Emparan}, 
will be addressed in a forthcoming paper.

\acknowledgments{One of us (C.R.O.) would like to thank 
the Kavli Institute for Theoretical Physics for generous support and for 
providing a wonderful environment during the completion of this work. 
Fruitful conversations with David Berenstein, Don Marolf, and Joe Polchinski 
are also acknowledged. This work was partially supported
by the National Science Foundation under Grant
No.\ 0602340  (H.E.C.)
and under Grants No.\ 0602301 and PHY05-51164 (C.R.O.), and
by the University of San Francisco Faculty Development Fund
(H.E.C.).}


\begin{thebibliography}{99}

\bibitem{Bekenstein_entropy}
J.D. Bekenstein,
{\em Black holes and entropy,
Phys. Rev.\/}  {\bf D 7}, 2333 (1973);
{\em Generalized second law of thermodynamics in black hole physics,
Phys. Rev.\/}  {\bf D 9} (1974) 3292.

\bibitem{BH-thermo_reviews}
For extensive reviews see,
R.M. Wald,
{\em The Thermodynamics of black holes,
Living Rev. Rel.\/} {\bf 4} (2001) 6 [{\sf gr-qc/9912119}];
\newline
T. Padmanabhan,
{\em Gravity and the thermodynamics of horizons,
Phys. Rept.\/} {\bf 406} (2005) 49
[{\sf gr-qc/0311036}].

\bibitem{thooft:85}
G. 't Hooft,
{\em On the Quantum structure of a black hole,
Nucl. Phys.\/} {\bf B 256} (1985) 727.

\bibitem{Frolov-Fursaev}
V.P. Frolov and D.V. Fursaev,
{\em Thermal fields, entropy, and black holes, 
Class. and Quant. Grav.\/} {\bf 15} (1998) 2041
[{\sf hep-th/9802010}]
and references therein.

\bibitem{holographic}
G. 't Hooft,
{\em Dimensional reduction in quantum gravity\/},
{\sf gr-qc/9310026};
\newline
L. Susskind,
{\em The world as a hologram,
J. Math. Phys\/} {\bf 36} (1995) 6377
[{\sf hep-th/9409089}];
\newline
R. Bousso,
{\em The holographic principle,
Rev. Mod. Phys.\/} {\bf 74} (2002) 825.
[{\sf hep-th/0203101}].

\bibitem{AdS/CFT}
J.M. Maldacena,
{\em The large N Limit of Superconformal Field Theories and
Supergravity,
Adv. Theor. Math. Phys.\/} {\bf 2} (1998)  231
[{\em Int. J. Theor. Phys.\/} {\bf 38} (1999) 1113]
[{\sf hep-th/9711200}].

\bibitem{entanglement}
 D. Kabat and M.J. Strassler,
{\em A comment on entropy and area, 
Phys. Lett.\/} {\bf B 329} (1994) 46
[{\sf hep-th/9401125}];
\newline
C.G. Callan Jr. and F. Wilczek,  
{\em On geometric entropy,
Phys. Lett.\/} {\bf B 333} (1994) 55
[{\sf hep-th/9401072}];
\newline
T. Jacobson, 
{\em A note on Hartle-Hawking vacua,
Phys. Rev.\/} {\bf D 50} (1994) 6031
\newline
[{\sf gr-qc/9407022}].

\bibitem{BH_thermo_CQM}
H.E. Camblong and C.R. Ord\'{o}\~{n}ez,
{\em Black hole thermodynamics from near-horizon conformal quantum mechanics,
Phys. Rev.\/}  {\bf D 71} (2005) 104029
[{\sf hep-th/0411008}].

\bibitem{semiclassical_BH_thermo}
H.E. Camblong and C.R. Ord\'{o}\~{n}ez,
{\em Semiclassical methods in curved spacetime and black hole thermodynamics,
Phys. Rev.\/} {\bf D 71} (2005) 124040
[{\sf hep-th/0412309}].

\bibitem{Fulling}
S.A. Fulling, 
{\em Aspects of quantum field theory in curved space-time,\/}
Cambridge University Press, Cambridge U.K. (1989). 

\bibitem{GR_Hawking}
 S.W. Hawking and W. Israel
{\em General relativity: an Einstein centenary survey,\/}
Cambridge University Press, Cambridge U.K. (1979).

\bibitem{Vassilevich}
 D.V. Vassilevich, 
{\em Heat kernel expansion: User's manual,
Phys. Rept.\/} {\bf 388} (2003) 279
[{\sf hep-th/0306138}].

\bibitem{Kac-drum}
 M. Kac, 
{\em Can one hear the shape of a drum,
%Am. Mathematical Monthly {\bf 73}, 1 (1966).
Am. Math. Month.\/} {\bf 73} (1966) 1.

\bibitem{optical_metric}
G.W. Gibbons and M.J. Perry,
{\em Black holes and thermal Green functions,
Proc. Roy. Soc. Lond.\/} {\bf A 358} (1978) 467;
\newline
J.S. Dowker and G. Kennedy, 
{\em Finite temperature and boundary effects in static space-times,
J. Phys.\/} {\bf A 11} (1978) 895;
\newline
G. Kennedy, R. Critchley, and J.S.  Dowker, 
{\em Finite temperature field theory with boundaries:
stress tensor and surface action renormalization,
Ann. Phys. (NY)\/} {\bf 125} (1980) 346.

\bibitem{deAlwis-Ohta}
S.P. de Alwis and N. Ohta,
{\em Thermodynamics of quantum fields in black hole backgrounds,
Phys. Rev.\/} {\bf D 52} (1995) 3529
[{\sf hep-th/9504033}].

\bibitem{Barbon-Emparan}
J.L.F.  Barb\'{o}n and R. Emparan,
{\em On quantum black hole entropy and Newton constant renormalization,
Phys. Rev.\/} {\bf D  52} (1995) 4527
[{\sf hep-th/9502155}].

\bibitem{forsyth:Liouville}
A.R. Forsyth,
{\em A treatise on differential equations\/}, 6th ed.,
Macmillan, London (1929).

\bibitem{susskind_uglum}
L. Susskind and J. Uglum,
{\em Black hole entropy in canonical quantum gravity and 
superstring theory,
Phys. Rev.\/} {\bf D 50} (1994) 2700
[{\sf hep-th/9401070}].

\bibitem{renormalization_Newton_confirmed}
S.N. Solodukhin,
{\em  On 'nongeometric' contribution to the entropy of black hole
due to quantum corrections,
Phys. Rev.\/} {\bf D 51} (1995) 618;
[{\sf hep-th/9408068}];
\newline
N.E. Mavromatos and E. Winstanley,
{\em Aspects of hairy black holes in spontaneously broken
Einstein Yang-Mills systems: stability analysis and entropy considerations,
Phys. Rev.\/} {\bf D 53} (1996) 3190
[{\sf hep-th/9510007}];
\newline
J.G. Demers, R. Lafrance, and R.C. Myers,
{\em Black hole entropy without brick walls,
Phys. Rev.\/} {\bf D 52} (1995) 2245
[{\sf gr-qc/9503003}];
\newline
D.V.  Fursaev and S.N. Solodukhin,
{\em On one loop renormalization of black hole entropy,
Phys. Lett.\/} {\bf B 365} (1996) 51
[{\sf hep-th/9412020}];
\newline
S.P. Kim, S.K. Kim, K.-S. Soh, and J.H. Yee,
{\em Remarks on renormalization of black hole entropy,
Int. J. Mod. Phys.\/} {\bf A 12} (1997) 5223
[{\sf gr-qc/9607019}];
\newline
T. Shimomura,
{\em On renormalization of black hole entropy and of coupling constants,
Phys. Lett.\/} {\bf B 480} (2000) 207;
\newline
E. Winstanley,
{\em Renormalized black hole entropy via the 'brick wall' method,
Phys. Rev.\/} {\bf D 63} (2001) 084013
[{\sf hep-th/0011176}].

\bibitem{carlip:near_horizon}
S. Carlip,
{\em Black hole entropy from conformal field theory in any dimension,
Phys. Rev. Lett.\/} {\bf 82} (1999) 2828
[{hep-th/9812013}];
{\em Entropy from conformal field theory at Killing horizons,
Class. and Quant. Grav.\/} {\bf 16} (1999) 3327
[{\sf gr-qc/9906126}];
{\em Black hole entropy from horizon conformal field theory, 
Nucl. Phys.\/} {\bf  88} {\em (Proc. Suppl.)}  (2000) 10
[{\sf gr-qc/9912118}];
\newline
{\em Near-horizon conformal symmetry and black hole entropy,
Phys. Rev. Lett.\/} {\bf 88} (2002) 241301
[{\sf gr-qc/0203001}];
{\em Horizon constraints and black hole entropy,
Class. and Quant. Grav.\/} {\bf 22} (2005) 1303
[{\sf hep-th/0408123}].

\bibitem{solodukhin:99}
S.N. Solodukhin,
{\em Conformal description of horizon's states,
Phys. Lett.\/} {\bf B 454} (1999) 213
[{\sf hep-th/9812056}].

\end{thebibliography}
\end{document}